\documentclass[11pt]{article}
\usepackage[letterpaper,margin=1in]{geometry}

\usepackage{amsmath}
\usepackage{graphicx}
\usepackage{amssymb}
\usepackage{amsthm}
\usepackage{booktabs}
\usepackage{enumitem}
\usepackage{xcolor}
\usepackage{tikz}
\usepackage{eso-pic}
\usepackage{url}
\usetikzlibrary{positioning,shapes.geometric,arrows.meta}

\providecommand{\citenamefont}[1]{#1}

\newtheorem{theorem}{Theorem}[section]

\newtheorem{definition}[theorem]{Definition}

\newcommand{\fito}{\textsc{fito}}
\newcommand{\oae}{\textsc{oae}}
\newcommand{\rdma}{\textsc{rdma}}

\title{\textbf{The Semantic Arrow of Time, Part~III:}\\[0.3em]{\large RDMA and the Completion Fallacy}}
\author{Paul Borrill \\ D\AE D\AE LUS}
\date{02026-FEB-27}

\usepackage{natbib}
\usepackage[hidelinks]{hyperref}

\begin{document}
\maketitle

\begin{center}
\large\itshape Why Delivery Guarantees Destroy Meaning
\end{center}
\vspace{1em}

\begin{abstract}
\noindent
This is the third of five papers comprising \emph{The Semantic Arrow
of Time}.  Part~I~\citep{borrill2026-partI} identified computing's
hidden arrow of time as semantic rather than thermodynamic and traced
the \fito{} category mistake from Eddington through Shannon and
Lamport.  Part~II~\citep{borrill2026-partII} presented the constructive
alternative: the \oae{} link state machine, Indefinite Logical
Timestamps, and the Slowdown Theorem's round-trip requirement.

This paper examines what happens when those principles are violated at
industrial scale.  Remote Direct Memory Access (\rdma{}) is the
highest-performance data movement technology in production use today,
deployed across Meta's 24,000-GPU Llama~3 training clusters, Google's
multi-tenant data centers, and Microsoft's Azure infrastructure.
\rdma{}'s core promise is simple: bypass the kernel, bypass the CPU,
and place data directly into remote memory at line rate.

We argue that this promise contains a category mistake of precisely the
form identified in Part~I.  \rdma{}'s completion semantics---the
signal that tells the sender ``the operation is done''---guarantee
\emph{placement} (data has been written to a remote NIC buffer) but
not \emph{commitment} (the data has been semantically integrated by the
receiving application).  We call this the \emph{completion fallacy}: the
conflation of delivery confirmation with semantic agreement.

We document the fallacy through seven temporal stages of an \rdma{}
Write operation, showing that the gap between stage~$T_4$ (completion
signal returned) and stage~$T_6$ (application semantic properties
satisfied) can be arbitrarily large.  We then trace the consequences
through four production-scale case studies: Meta's RoCE fabric at
32,000 GPUs, Google's 1RMA redesign, Microsoft's DCQCN
interoperability failures, and the SDR-\rdma{} partial completion
problem.

The paper concludes with a comparative semantic analysis showing that
CXL~3.0, NVLink, and UALink each address parts of the completion
fallacy but none eliminates it entirely.  Only a protocol architecture
with a mandatory reflecting phase---as specified in Part~II---can
close the gap between delivery and commitment.
\end{abstract}

\section[Introduction]{Introduction: The Fastest Way to Corrupt Data}
\label{sec:intro}

Remote Direct Memory Access is fast.  It bypasses the operating system
kernel, avoids CPU intervention on the data path, and writes directly
into application memory at rates exceeding 400~Gb/s on current
hardware.  For the AI training workloads that now dominate hyperscale
data centers, \rdma{} is not optional---it is the only technology that
can move gradient tensors between thousands of GPUs within the
synchronization budget of a training step.

But speed is not the same as correctness.  The central claim of this
paper is that \rdma{}'s completion semantics implement the \fito{}
category mistake at the hardware level: they treat the forward
delivery of data as sufficient evidence that communication has occurred.
In the terminology of Part~II, \rdma{} conflates \emph{placement} with
\emph{commitment}---it signals ``done'' at a point in the temporal
sequence where the semantic arrow has not yet been established.

The consequence is not occasional data loss.  It is \emph{systematic
semantic corruption}: data structures that parse correctly but contain
inconsistent state, completion signals that report success while
multi-field invariants are violated, and collective operations that
diverge silently because individual transfers completed before
visibility was guaranteed.

The remainder of this paper is organized as follows.
Section~\ref{sec:promise} examines \rdma{}'s architectural promise and
the engineering assumptions behind it.
Section~\ref{sec:seven-stages} decomposes an \rdma{} Write into seven
temporal stages, showing exactly where the completion fallacy occurs.
Section~\ref{sec:atomicity-gap} analyzes the 8-byte atomicity boundary
and its consequences for real data structures.
Section~\ref{sec:production} presents four production-scale case studies.
Section~\ref{sec:sdc} connects the completion fallacy to the OCP Silent
Data Corruption initiative.
Section~\ref{sec:comparison} provides a comparative analysis of CXL,
NVLink, and UALink.
Section~\ref{sec:summary} summarizes and previews Part~IV.

\section[\rdma{}'s Promise]{\rdma{}'s Architectural Promise}
\label{sec:promise}

\rdma{} was designed to solve a specific problem: the overhead of
traditional networking stacks.  In a conventional TCP/IP path, a
network write requires a system call (kernel transition), a copy from
user space to kernel buffer, protocol processing (TCP segmentation,
IP encapsulation), a DMA transfer to the NIC, transmission, and the
reverse path at the receiver.  Each stage adds latency and consumes
CPU cycles.%

\rdma{} eliminates most of this path.  The application posts a Work
Request to a Queue Pair (QP) on the NIC; the NIC reads the data
directly from registered application memory via DMA; the NIC
handles segmentation, transmission, and reassembly; and at the
receiver, the NIC writes directly into the remote application's
registered memory without involving the remote CPU.  The only
software involvement is posting the request and polling for
completion.

Three classes of operations are provided:

\begin{description}[leftmargin=2cm]
  \item[\textsc{send/recv}:] Two-sided operations requiring receiver
    cooperation.  The receiver posts a Receive buffer; the sender
    transmits into it.  Both sides are involved.
  \item[\textsc{write}:] One-sided operation.  The sender specifies a
    remote address and writes data there without any involvement of
    the remote CPU.  The remote application is not notified.
  \item[\textsc{atomic}:] Compare-and-Swap (CAS) and Fetch-and-Add
    (FAA) on remote memory.  Atomic operations are limited to 8 bytes
    and execute at the remote NIC, not the remote CPU.
\end{description}

One-sided operations are the purest expression of \fito{} at the
hardware level.  They model communication as Shannon modeled his
channel: sender transmits, channel delivers, receiver is passive.
The return path---the acknowledgment---is handled at the transport
level (reliable connection mode) but carries no semantic content.
It confirms that the NIC accepted the data, not that the application
understood it.

\section[Seven Stages]{The Seven Temporal Stages of an \rdma{} Write}
\label{sec:seven-stages}

To make the completion fallacy precise, we decompose an \rdma{} Write
into its temporal stages.  Each stage represents a distinct point in
the lifecycle of the operation; the completion fallacy is the
assumption that one of these stages implies a later one.

\begin{definition}[Seven Stages]
\label{def:seven-stages}
\mbox{}
\begin{description}[leftmargin=1.5cm, font=\normalfont\bfseries]
  \item[$T_0$:] \textbf{Submission.}  The application posts a Write
    Work Request to the Send Queue of a Queue Pair.  The NIC has not
    yet touched the data.
  \item[$T_1$:] \textbf{Placement.}  The NIC reads the data from
    registered application memory via DMA and stages it in the NIC's
    internal buffers.  The data is now ``placed'' in the network
    subsystem.
  \item[$T_2$:] \textbf{Transmission.}  The NIC segments the data into
    network packets, adds headers, and transmits them onto the wire
    (InfiniBand) or Ethernet fabric (RoCE).
  \item[$T_3$:] \textbf{Remote Placement.}  The remote NIC receives
    packets and writes them directly into the specified remote memory
    address via DMA.  The data is now in remote memory but may reside
    in posted write buffers, not yet visible through the cache
    hierarchy.
  \item[$T_4$:] \textbf{Completion.}  The sending NIC receives a
    transport-level acknowledgment and posts a Completion Queue Entry
    (CQE) to the sender's Completion Queue.  The application polls
    or is notified that the Write ``succeeded.''
  \item[$T_5$:] \textbf{Visibility.}  The data becomes visible to the
    remote CPU through cache coherence mechanisms.  This may require
    the remote CPU to execute a memory fence or read the location,
    triggering a cache fill.  $T_5$ can be significantly later than
    $T_4$.
  \item[$T_6$:] \textbf{Semantic Agreement.}  The remote application
    reads the data, validates it against application invariants, and
    integrates it into its semantic state.  Multi-field consistency,
    version compatibility, and causal ordering are established.
\end{description}
\end{definition}

The completion fallacy is the assumption that $T_4 \Rightarrow T_6$:
that the completion signal guarantees semantic agreement.  In fact,
the gap between $T_4$ and $T_6$ can be arbitrarily large, and the
operations between them---cache coherence, memory ordering, invariant
validation, causal integration---are entirely outside \rdma{}'s
purview.%

In the \oae{} framework of Part~II, the stages map cleanly:

\begin{center}
\small
\begin{tabular}{lll}
\toprule
\textbf{Stage} & \textbf{\rdma{} status} & \textbf{\oae{} equivalent} \\
\midrule
$T_0$--$T_1$ & Submitted & \textsc{idle} $\to$ \textsc{tentative} \\
$T_2$--$T_3$ & In flight & \textsc{tentative} \\
$T_4$        & Completed & \textsc{tentative} (still!) \\
$T_5$        & Visible   & \textsc{tentative} (still!) \\
$T_6$        & Agreed    & \textsc{committed} \\
\midrule
& \textsc{reflecting} & \emph{absent from \rdma{}} \\
\bottomrule
\end{tabular}
\end{center}

The table makes the fallacy visible: \rdma{} reports ``completed'' at
a point that \oae{} would still classify as \textsc{tentative}.
There is no reflecting phase; there is no mechanism for the receiver to
confirm that it has semantically processed the data.  The completion
signal is a transport-level acknowledgment masquerading as an
application-level commitment.

\section[The 8-Byte Boundary]{The Atomicity Gap: The 8-Byte Boundary}
\label{sec:atomicity-gap}

\rdma{} provides two atomic operations: Compare-and-Swap and
Fetch-and-Add.  Both are limited to 8 bytes (64 bits) and execute at
the remote NIC, not the remote CPU.  This creates what we call the
\emph{atomicity gap}: the space between what \rdma{} guarantees
atomically (8 bytes) and what applications require atomically
(arbitrary data structures).

Consider a distributed hash table entry:
\begin{center}
\small
\begin{tabular}{lrl}
\toprule
\textbf{Field} & \textbf{Size} & \textbf{Content} \\
\midrule
\texttt{version} & 8 bytes & monotonic counter \\
\texttt{key}     & 32 bytes & hash key \\
\texttt{value}   & 256 bytes & payload \\
\texttt{status}  & 8 bytes & \{valid, deleted, locked\} \\
\bottomrule
\end{tabular}
\end{center}

This 304-byte structure spans at least 5 cache lines (at 64 bytes each).
An \rdma{} Write that updates this entry is \emph{not} atomic: a
concurrent reader may observe the new \texttt{version} with the old
\texttt{value}, or the new \texttt{value} with the old \texttt{status}.
The entry parses correctly---every field is individually well-formed---but
the combination is semantically invalid.  This is the definition of
semantic corruption: syntactically correct, semantically meaningless.%

The standard mitigation is FaRM's versioned approach~\citep{farm2015}:
embed a version number in the last 8 bytes of each cache line, use
\rdma{} Write to update the data, and use a final \rdma{} Atomic CAS
to commit the version.  The reader checks the version before and after
reading; if the versions differ, the read is retried.

This is timeout-and-retry (\textsc{tar}) at the cache-line level---precisely
the \fito{} pattern that Part~I identified as the root of distributed
systems pathology.  FaRM does not solve the atomicity gap; it papers
over it with forward-only retries that assume lost writes and hope for
eventual success.  The semantic arrow is not established; it is
approximated through probabilistic convergence.

\section[Production Evidence]{Production-Scale Evidence}
\label{sec:production}

The completion fallacy is not a theoretical concern.  It manifests at
every scale of \rdma{} deployment, from single-machine NUMA domains to
hyperscale training clusters.  We present four case studies.

\subsection{Case Study 1: Meta's Llama~3 at 24,000 GPUs}

Gangidi et al.~\citep{gangidi2024} report Meta's deployment of two
24,000-GPU clusters---one RoCE, one InfiniBand---for Llama~3 training.
The RoCE cluster exposed several completion-fallacy pathologies:

\paragraph{Congestion cascades.}
RoCE relies on Priority Flow Control (PFC) to implement lossless
delivery.  When congestion builds, PFC pause frames propagate
backward through the fabric, stalling senders.  But PFC operates at
the link level, not the semantic level: it pauses \emph{all} traffic
on a priority class, including traffic that is not contributing to
the congestion.  The result is head-of-line blocking that delays
completion signals for operations that have already been placed in
remote memory---widening the $T_3$-to-$T_4$ gap and making the
$T_4$-to-$T_6$ gap unpredictable.%

\paragraph{Low-entropy traffic.}
AI training produces collective communication patterns (all-reduce,
all-gather, reduce-scatter) that generate a small number of large,
synchronized flows.  ECMP hash-based load balancing assumes
high-entropy traffic---many small, independent flows.  When traffic
entropy is low, ECMP degenerates: multiple flows hash to the same
path, creating persistent hotspots.  Meta reports that QP-based
flow scaling (opening multiple Queue Pairs per connection to increase
hash entropy) improved throughput by 40\%.

This is a category-mistake symptom: the network fabric assumes that
traffic is statistically independent (\fito{} at the flow level),
but AI collective operations are causally correlated.  The fabric's
model of traffic does not match the traffic's semantic structure.

\paragraph{Transport livelock.}
Even rare packet losses---below $10^{-6}$---consistently degraded
performance because \rdma{}'s reliable transport retransmits entire
messages upon detecting any loss.  A single dropped packet in a 1~GiB
transfer triggers retransmission of the entire 1~GiB.  The completion
signal for the original transfer is delayed or never arrives; the
application stalls waiting for a completion that semantic agreement
would have resolved more efficiently through partial acknowledgment.

\subsection{Case Study 2: Google's 1RMA Redesign}

Singhvi et al.~\citep{singhvi2020} at Google concluded that
standard \rdma{} is ``ill-suited to multi-tenant datacenters'' and
designed 1RMA from scratch.  The key insight was that \rdma{}'s
connection-oriented model (Queue Pairs bound to specific remote
endpoints) creates isolation, fairness, and security problems at
cloud scale.

1RMA treats each \rdma{} operation as an independent, connectionless
request.  Software provides congestion control, failure recovery, and
inter-operation ordering---functions that \rdma{} delegates to hardware
QP state machines.

In the framework of this series, Google's redesign is a partial
correction of the \fito{} assumption: by making each operation
independent, 1RMA avoids the assumption that a sequence of operations
to the same endpoint should be ordered by their submission times.
But 1RMA retains the completion fallacy: the NIC still signals
``done'' at $T_4$, and the gap to $T_6$ is still the application's
problem.

\subsection{Case Study 3: Microsoft DCQCN Interoperability}

Microsoft's Azure \rdma{} deployment~\citep{microsoft-rdma2023}
exposed a subtler form of the completion fallacy: interoperability
failures between NIC generations.

When Gen2/Gen3 NICs (hardware-based congestion control) communicated
with Gen1 NICs (firmware-based), the completion semantics diverged:
Gen2/Gen3 NICs generated Congestion Notification Packets (CNPs) at a
rate that overwhelmed Gen1's firmware processing, causing excessive
rate reduction.  In the reverse direction, Gen1's timer-based CNP
coalescing produced bursts that caused cache misses in Gen2/Gen3's
receiver pipeline.

The result was not packet loss---the transport layer continued to
report successful completions.  It was \emph{performance collapse}:
completion signals arrived, but the effective throughput dropped by
an order of magnitude because the congestion control mechanisms were
fighting each other.  Completions succeeded; meaning (in this case,
the intended throughput) was destroyed.%

\subsection{Case Study 4: SDR-\rdma{} and Partial Completion}

Khalilov et al.~\citep{khalilov2025} identified a particularly
damaging form of the completion fallacy in Unreliable Connected (UC)
mode.  When a single 4~KiB packet is lost during a 1~GiB Write, the
\emph{entire} 1~GiB operation is marked as lost.  The completion
signal reports failure for the whole message, even though 99.9996\%
of the data arrived successfully.

SDR-\rdma{} fixes this with a partial completion API: a bitmap
indicates which chunks were dropped, enabling selective
retransmission.  This is a significant engineering improvement, but
it reveals the depth of the problem: \rdma{}'s completion semantics
are so coarse that a single lost packet invalidates a gigabyte of
successfully placed data.  The completion signal's binary
(success/failure) model cannot represent the actual state of the
transfer, which is partially placed and partially absent.

In the \oae{} framework, this is a failure of the
\textsc{reflecting} phase: the receiver cannot communicate
\emph{what it actually received} back to the sender.  The
completion signal is a 1-bit response (success or failure) where a
semantic digest---a structured description of the receiver's
state---is needed.

\section[Silent Data Corruption]{The Completion Fallacy and Silent Data Corruption}
\label{sec:sdc}

The Open Compute Project's Silent Data Corruption (SDC)
initiative~\citep{ocp-sdc2024}, launched in October 2022 with founding
members AMD, ARM, Google, Intel, Meta, Microsoft, and NVIDIA,
addresses hardware-induced data corruption that escapes error detection
mechanisms.  SDC is the ultimate expression of the completion fallacy:
the hardware executes an operation, reports success, and produces
corrupted output.

SDC and the completion fallacy interact in a particularly dangerous
way in \rdma{} systems.  Consider the temporal sequence:

\begin{enumerate}[leftmargin=1.2cm]
  \item An \rdma{} Write places data in remote memory ($T_3$).
  \item A hardware fault (manufacturing defect, aging, cosmic ray)
    corrupts one or more bits during placement, in the NIC buffer,
    or in the DMA path.
  \item The sending NIC receives a transport-level ACK ($T_4$) and
    reports successful completion.
  \item The corruption is undetected by the NIC's error detection
    (CRC covers the transport frame, not the DMA path; ECC covers
    DRAM, not NIC buffers).
  \item The application reads the corrupted data ($T_6$) and treats
    it as valid---the completion signal said the Write succeeded.
\end{enumerate}

The completion signal is not merely insufficient---it is
\emph{actively misleading}.  It tells the application that the
operation succeeded when the data is corrupt.  The application has no
mechanism within \rdma{}'s semantics to discover the corruption; it
must rely on application-level checksums or end-to-end integrity
verification---mechanisms that \rdma{}'s ``zero-copy, kernel-bypass''
architecture was designed to eliminate.%

The OCP's 2024 AI-specific SDC white paper~\citep{ocp-sdc2024}
documents the consequences for AI training: silent corruption of
gradient tensors during all-reduce can propagate through the training
loop, producing models with degraded accuracy that are
indistinguishable from undertrained models.  The completion fallacy
means that the training framework believes every gradient exchange
succeeded; the semantic corruption is invisible until the model is
evaluated, potentially weeks later.

\section[Comparative Analysis]{Comparative Analysis: CXL, NVLink, UALink}
\label{sec:comparison}

Each major interconnect technology addresses aspects of the completion
fallacy, but none eliminates it entirely.

\subsection{CXL~3.0: Coherence Without Commitment}

CXL~3.0's back-invalidation mechanism~\citep{cxl30} introduces
hardware-enforced cache coherence between hosts and CXL devices.
When a device modifies a cache line, it issues a Back-Invalidate
snoop (BISnp) to all host caches that may hold stale copies.
The Device Coherency Engine (DCOH) tracks host cache state using
a MESI-like protocol.

This addresses the $T_4$-to-$T_5$ gap: after a CXL write completes,
cache coherence ensures that subsequent reads by any host see the
updated data.  But CXL's coherence operates at cache-line
granularity (64--128 bytes).  A multi-cache-line update is not
atomic: readers may observe a mix of old and new cache lines.  The
$T_5$-to-$T_6$ gap---from visibility to semantic agreement---remains
entirely the application's responsibility.

CXL's Unordered I/O (UIO) mechanism further complicates matters.
Virtual channels VC1--VC7 carry traffic that is not subject to
strict ordering rules, enabling higher throughput but breaking the
assumption that operations to the same address region complete in
submission order.  The completion signal for a UIO operation
guarantees delivery to the fabric, not visibility at the
destination---a completion fallacy within the fabric itself.

\subsection{NVLink: Signals Without Semantics}

NVLink's signal-based ordering model provides the strongest
completion guarantees of any production interconnect.  A signal
operation atomically updates a remote counter; when the signal
arrives, all preceding put operations on the same context are
guaranteed visible at the destination.

This eliminates the $T_4$-to-$T_5$ gap for operations within a
single context: visibility is guaranteed at signal completion.
But the signal carries no semantic content.  It confirms that data
is visible; it does not confirm that the data is correct,
consistent, or interpretable.  If the sender transmits a
multi-field update and the receiver interprets the fields
differently (due to version skew, schema evolution, or
endianness mismatch), the divergence is undetectable.  The
signal guarantees delivery; it does not guarantee understanding.%

NVLink's scope also limits its applicability.  It is an intra-node
interconnect (extending to small clusters of 8 nodes with NVLink
Switch); it does not extend to the data center scale where \rdma{}'s
completion fallacy is most damaging.

\subsection{UALink: Memory Semantics Without Reflection}

The UALink consortium's interconnect adopts memory-semantic
operations (load/store) rather than \rdma{}'s verb-based model
(Send/Write/Read/Atomic).  Operations use fixed 640-byte flits,
eliminating Ethernet framing overhead and the packet aggregation
that introduces variable latency.

UALink's memory semantics are closer to \oae{}'s model than
\rdma{}'s: a load/store operation implies that the remote memory
system participates in the operation, not merely the remote NIC.
But UALink is designed for scale-up fabric (intra-pod, intra-rack)
and inherits the cache coherence model of its member companies'
existing architectures.  It has no published specification for
multi-flit atomic guarantees, no reflecting mechanism, and no
semantic digest.  It is faster and cleaner than \rdma{}, but it
remains a \fito{} technology: data moves forward, and the return
path carries only transport-level status.

\subsection{Summary Table}

\begin{table}[h]
\small
\begin{center}
\begin{tabular}{lccccc}
\toprule
\textbf{Gap} & \textbf{\rdma{}} & \textbf{CXL} & \textbf{NVLink}
  & \textbf{UALink} & \textbf{\oae{}} \\
\midrule
$T_4 \to T_5$ (visibility) & open & closed & closed & unknown & closed \\
$T_5 \to T_6$ (semantics) & open & open & open & open & closed \\
Atomicity boundary & 8~B & 64--128~B & 128~B & 640~B & transaction \\
Reflecting phase & absent & absent & absent & absent & mandatory \\
SDC detection & none & none & none & none & reflection mismatch \\
\bottomrule
\end{tabular}
\end{center}
\caption{Temporal gap analysis across interconnect technologies.
Only \oae{} closes the $T_5$-to-$T_6$ gap through its mandatory
reflecting phase.}
\label{tab:gaps}
\end{table}

\section[Summary]{Summary and Preview of Part~IV}
\label{sec:summary}

This paper has argued that \rdma{}'s completion semantics implement the
\fito{} category mistake at the hardware level.  The completion
signal---the CQE posted to the sender's Completion Queue---reports
success at temporal stage~$T_4$, when the semantic arrow has not yet
been established.  The gap between $T_4$ (completion) and $T_6$
(semantic agreement) is the space in which meaning is destroyed:
multi-field invariants are violated, silent data corruption propagates
undetected, and completion signals actively mislead applications about
the state of their transactions.

The four case studies demonstrate that this is not a theoretical
concern but an operational reality at the largest scales of deployment.
Meta's 24,000-GPU clusters, Google's multi-tenant data centers,
Microsoft's heterogeneous NIC deployments, and ETH Zurich's SDR-\rdma{}
work all document the same pattern: \rdma{} completes; meaning is lost.

CXL, NVLink, and UALink each address parts of the problem---cache
coherence, signal-based ordering, memory semantics---but none provides
the reflecting phase that closes the $T_5$-to-$T_6$ gap.  The
completion fallacy persists across the entire interconnect landscape
because it is embedded in the \fito{} assumption that every technology
shares: information flows forward, and the return path is auxiliary.

Part~IV~\citep{borrill2026-partIV} traces the consequences of the
completion fallacy beyond the data center, into the systems that
people use every day: file synchronization that silently deletes
content, email that reorders causality, and human memory that
reconstructs rather than replays---each a system where the semantic
arrow is routinely violated, and where the consequences are borne not
by NICs and GPUs but by the people who depend on them.


\end{document}